\documentclass[12pt,aps,prb,preprint]{revtex4}   
\usepackage{amsmath}   
\usepackage{amsfonts}  
\usepackage{amssymb}

\begin{document}

\title{Spin and Statistics in Classical Mechanics}

\author{J. A. Morgan}
\affiliation{The Aerospace Corporation\\P. O. Box 92957\\Los Angeles, CA 90009}
\email{john.a.morgan@aero.org}   

\begin{abstract}
The spin-statistics connection is obtained for classical point particles.  The 
connection holds within pseudomechanics, a theory of particle motion that extends classical 
physics to include anticommuting Grassmann variables, and which exhibits classical analogs 
of both spin and statistics.  Classical realizations of Lie groups can be constructed in a 
canonical formalism generalized to include Grassmann variables.  The theory of irreducible 
canonical realizations of the Poincar\'{e} group is developed in this framework, with particular
emphasis on the rotation subgroup.  The behavior of irreducible realizations under time inversion 
and charge conjugation is obtained.  The requirement that the Lagrangian retain its form under 
the combined operation $CT$ leads directly to the spin-statistics connection, by an adaptation of 
Schwinger's 1951 proof to irreducible canonical realizations of the Poincar\'{e} group of 
spin $j$:  Generalized spin coordinates and momenta satisfy fundamental Poisson bracket relations 
for 2$j$=even, and fundamental Poisson antibracket relations for 2$j$=odd.
\end{abstract}

\maketitle

\section{Introductory}

"In conclusion we wish to state, that according to our opinion, the connection between spin 
and statistics is one of the most important applications of the special relativity theory."  
W. Pauli, 1940 

The spin-statistics connection was originally 
established as a theorem in relativistic quantum field theory.  It has acquired a 
formidable reputation, not just from the undeniable difficulty of the papers which originally 
established the connection in field theory, but also from 
the vicissitudes of later attempts to simplify and streamline proofs of the 
theorem.  Over time, these latter efforts have increasingly come to
concentrate on the quantum-mechanical at the expense of the relativistic.  In contrast, this 
paper presents an example of a \emph{classical} dynamical system that obeys a spin-statistics 
relation as a consequence of local Poincar\'{e} invariance.  
This project is very much in line 
with the sentiment, quoted above, expressed more than sixty years ago by Pauli in his landmark 
paper.~\cite{Pauli1940}  It does go-somewhat-against the grain 
of much of the subesquent research on spin and statistics.  

Developments in the intervening decades 
have fallen, roughly speaking, under three heads:  First, the basic result has seen a deepening 
and strengthening considered as a theorem in relativistic quantum field 
theory.\cite{Pauli1946}$^{-}$~\cite{Good1958}  A collection of the significant papers marking 
this evolution has been reprinted in the critical retrospective volume by Duck and 
Sudarshan~\cite{DuckS1997}(\emph{vide.} also the paper by Greenberg.~\cite{G1998})    

Second, while special relativity in the form of local Poincar\'{e} symmetry is a sufficient 
condition for a local quantum field theory to obey the spin-statistics connection, numerous 
investigations have sought a weaker set of necessary conditions.  Topological considerations in 
a non-relativistic setting predominate 
in these studies.~\cite{Tsch1989}$^{-}$~\cite{Balach1993}  In the course of this search, 
the spin-statistics connection has been extended to settings sometimes far removed from 
relativistic field theory. Topological 
theorems have been obtained for strings~\cite{Balach1992a} and solitons on two-dimensional 
surfaces~\cite{Balach1991}.   A notable aspect of the topological theorems is that, 
while they make no explicit use of relativity or of quantum field theory, they do require the 
existence of antiparticles.~\cite{DS2001}  

Third, since the axiomatic proofs of Burgoyne~\cite{B1958} and L\"{u}ders and Zumino~\cite{LZ1958} 
there has been continuing interest in finding simple and elementary proofs of the spin-statistics 
relation.  All demonstrations of the spin-statistics theorem in quantum theory amount to 
\emph{proving}-with greater, or lesser, amounts of travail-what can be \emph{stated} very 
simply: The operation of exchanging the position of two identical particles is equivalent to the 
rotation of one of them by 2$\pi$.~\cite{footnote1}  
Feynman,~\cite{Feynman1949}$^{-}$\cite{Feynman1987} 
Neuenschwander,~\cite{Neuen1994} Duck and Sudarshan~\cite{DuckS1998} and others have addressed 
this goal explicitly.  It is at the least implicit in the long search for topological theorems by 
Balachandran \emph{et al.}~\cite{Balach1990}$^{-}$~\cite{Balach1991} and, from a 
different perspective, the proof of Berry and Robbins~\cite{BR1997}.

A common thread in both the search for necessary conditions and for simplified proofs is 
retreat both from field theory and from explicit reliance upon relativistic formalism.  
Despite this last, it appears that any proof of the spin-statistics connection requires assumptions 
traceable to local Poincar\'{e} symmetry.  All proofs depend upon rotational symmetry.  
The topological proofs, in addition, require antiparticles.~\cite{footnote2}  But, whatever 
the ultimate status of relativistic assumptions in the topological theorems, they demonstrate 
that the necessary conditions for the spin-statistics relation can be weak indeed.  

This paper addresses a variation on that observation: It is not necessary for a physical 
system to be quantum-mechanical in order to obey a spin-statistics relation.  For this 
statement to make sense, one needs classical notions of spin, and of statistics, both of 
which exist, and appear in the following.  That the spin-statistics connection is not 
intrinsically a quantum mechanical relation should not come entirely as a surprise.  The 
early topological theorem for kinks~\cite{RF1968} by Rubinstein and Finkelstein invokes few 
assumptions of an overtly quantum nature, while Mickelsson~\cite{Mick1984} explicitly proved a 
topological theorem valid for classical as well as quantum systems.  

Apart from Mickelsson's paper, little attention seems to have been devoted to classical analogs 
of the spin-statistics connection.  There has been great interest, on the other hand, in 
classical descriptions of spin, and of spinning 
electrons.~\cite{Corben1961}$^{-}$~\cite{ChoKim1994}  In particular, a classical theory of 
spinning particles constructed from 
anticommuting Grassmann variables finds practical use in constructing path integral formulations 
of supersymmetry.~\cite{BM1977}$^{,}$~\cite{GalvaoTeitelboim1980}  This formulation of classical 
mechanics of anticommuting dynamical variables has odd features, and has come to be labeled 
"pseudoclassical" mechanics in consequence.  However, the pseudoclassical theories prove 
suitable for exhibiting a classical spin-statistics connection.

While there is no attempt in this paper to evade explicit reliance upon relativistic symmetry 
in the form of local Poincar\'{e} invariance, the reasoning, in common with most proofs in recent 
decades ~\cite{Weinberg1964}$^{,}$~\cite{BR1997} is not, in fact, all that 
relativistic in detail.  The uses made of Poincar\'{e} invariance amount to two:  (1) The 
properties of the rotational subgroup, specifically the properties of irreducible canonical 
realizations of spin degrees of freedom, and (2) The combination of the discrete symmetries 
time-reversal $T$ and charge conjugation $C$.

Our starting point is a review of the properties of Grassmann variables, and the extension of 
the canonical formalism to classical Grassmann variables.  The properties of anticommuting 
Grassmann variables supply a classical equivalent of fermionic exchange symmetry.  Classical 
Lagrangians constructed from these variables are the simplest models of classical half-integral
spin.  

Next, canonical 
realizations of continuous (Lie) symmetry groups are described.  The infinitesimal generators of 
transformations produced by a Lie group form a finite algebra, the Lie algebra of the 
group~\cite{Geroch1985}$^{,}$~\cite{Tung1985}.  The algebra is expressed
in terms of bracket relations.  The Lie bracket used in unitary representations of symmetries
acting on quantum-mechanical Hilbert spaces is the familiar commutator (or anticommutator).  In 
the canonical formalism, the 
place of quantum-mechanical irreducible unitary representations is taken by that of irreducible 
canonical realizations.  Commutators are replaced by the equivalent Poisson brackets.  Canonical 
equivalents exist for the entire apparatus of unitary representations in a Hilbert space, 
including ladder and Casimir operators.  

The theory of canonical realizations of the Poincar\'{e} group for massive particles is developed, 
with emphasis on the rotation subgroup and the properties of dynamical variables corresponding 
to a definite spin. The classical analog of an elementary particle then becomes an irreducible 
canonical realization of the Poincar\'{e} group.~\cite{SMukunda1974}  The (anticanonical) 
time inversion operation $T$ is defined and its action on irreducible canonical realizations is 
exhibited.  A classical analog of charge-conjugation $C$ is introduced.  The composition of $T$ and 
$C$, called strong time inversion, is also sometimes called Weyl time inversion.  

Finally, following Schwinger, it is shown that invariance of the Lagrangian under strong time 
inversion implies the spin-statistics connection.  

Notation: Except as otherwise indicated in the text, lower case Greek letters can be either 
even or odd Grassmann variables.  When it is desirable to distinguish the even variables, 
these will sometimes be lower case Latin letters.  An asterisk denotes complex conjugation.  
A spacelike convention is assumed for the Minkowski metric $\eta =$ diag(-1,1,1,1).  The 
summation convention applies to repeated indices. 

\section{Grassmann variables in classical mechanics}

\subsection{Grassmann variables}

The classical analogs of quantum-mechanical fermionic and bosonic exchange symmetry are found in 
the properties of Grassmann variables.~\cite{Berezin1966}$^{,}$~\cite{Swanson1992}  The
creation and annihilation operators of field theory are familiar examples of Grassmann variables 
in quantum theory.  Even Grassmann 
variables commute, and correspond to the usual bosonic $c$-number variables of classical 
mechanics.  Even and odd variables, in either order, commute.  A set of $n$ odd real Grassmann 
variables obeys anticommutation relations
\begin{equation}
\xi_{\mu}\xi_{\nu}+\xi_{\nu}\xi_{\mu}=0
\end{equation}
for $\mu,\nu<=n$.  Thus, 
\begin{equation}
\xi^{2}_{\mu}=0.
\end{equation}
Anticommutativity of odd classical Grassmann variables is a classical form of the exclusion 
principle.

Differentiation on Grassmann variables can act from the right or the left.  The sign of the 
derivative of a product, for example, depends upon which derivative is used.  Left 
differentiation, in accord with the convention in Berezin,~\cite{Berezin1966} is assumed in 
the following. 

Determining the behavior of Grassman variables under the time-reversal transformation used in
Sections \ref{T} and \ref{SST} requires their properties under complex conjugation.  
Given two real odd variables $\xi^{R}$ and $\xi^{I}$ a complex Grassmann variable is defined by
\begin{equation}
\xi=\xi^{R}+i\xi^{I}
\end{equation}
with modulus squared
\begin{equation}
\xi^{*}\xi=i(\xi^{R}\xi^{I}-\xi^{I}\xi^{R})=2i\xi^{R}\xi^{I}.
\end{equation}
One desires this quantity to be real.  Equating it to its complex conjugate,
\begin{eqnarray}
(\xi^{*}\xi)^{*}=-i((\xi^{R}\xi^{I})^{*}-(\xi^{I}\xi^{R})^{*}) \\
=2i(\xi^{I}\xi^{R})^{*}=2i\xi^{R}\xi^{I}. \nonumber
\end{eqnarray}
Since this relation must hold for an arbitrary complex Grassmann variable, it must be that for 
any two real Grassmann variables $\eta$ and $\xi$,
\begin{equation}
(\eta\xi)^{*}=\xi\eta.
\end{equation}
It follows~\cite{Swanson1992a} that the complex conjugate of a product of complex Grassmann 
variables is
\begin{equation}
(\xi_{1}\xi_{1}\cdots\xi_{n})^{*}=\xi_{n}^{*}\cdots\xi_{2}^{*}\xi_{1}^{*}. \label{eq:cconj}
\end{equation}

\subsection{Extension of Canonical Formalism to Grassmann variables} 

We consider the canonical formalism for massive particles only.~\cite{Boltznote}  
Let $q_{i}$, $p_{i}$, 
$i=1,m$ be coordinates and momenta of even variables, and $\xi_{\alpha}$, $\pi_{\alpha}$, 
$\alpha=1,n$ be coordinates and momenta of odd variables. Given a Lagrangian 
\begin{equation}
L=L(q_{i},p_{i},\xi_{\alpha},\pi_{\alpha})
\end{equation}
the generalized Hamiltonian is given by
\begin{equation}
H=q_{i}p^{i}+\xi_{\alpha}\pi^{\alpha}-L
\end{equation}
and Hamiltons' equations become
\begin{equation}
\dot{p}^{i}=-{\partial H \over \partial q_{i}} \;\;\;\; 
\dot{q}_{i}={\partial H \over \partial p^{i}}
\end{equation}
\begin{equation}
\dot{\pi}^{\alpha}=-{\partial H \over \partial \xi_{\alpha}} \;\;\;\; 
\dot{\xi}_{i}=-{\partial H \over \partial \pi^{\alpha}}.
\end{equation}
The momenta are defined by
\begin{equation}
p^{i}={\partial L \over \partial \dot{q}_{i}} \;\;\;\; 
\pi^{i}={\partial L \over \partial \dot{\xi}_{i}}.
\end{equation}
In a theory containing even and odd Grassmann variables, the definition of the Poisson bracket 
generalizes.  The Poisson bracket of two even variables $f, g$ is given 
by~\cite{Casalbuoni1976}$^{,}$~\cite{Casalbuoni1976a}
\begin{eqnarray}
[f,g]=\left \{ {\partial f \over \partial q_{i}} {\partial g \over \partial p^{i}} -
{\partial g \over \partial q_{i}} {\partial f \over \partial p^{i}} \right \} +
\left \{ {\partial f \over \partial \xi_{\alpha}} {\partial g \over \partial \pi^{\alpha}} -
{\partial g \over \partial \xi_{\alpha}} {\partial f \over \partial \pi^{\alpha}} \right \} \\
=-[g,f]. \nonumber
\end{eqnarray}
The bracket of two odd variables $\theta,\pi$ is given by
\begin{eqnarray}
[\theta,\psi]=\left \{ {\partial \theta \over \partial q_{i}} {\partial \psi \over 
\partial p^{i}} +
{\partial \psi \over \partial q_{i}} {\partial \theta \over \partial p^{i}} \right \} -
\left \{ {\partial \theta \over \partial \xi_{\alpha}} {\partial \psi \over \partial 
\pi^{\alpha}} +
{\partial\psi \over \partial \xi_{\alpha}} {\partial \theta \over \partial \pi^{\alpha}} 
\right \} \\
=[\psi,\theta] \nonumber
\end{eqnarray}
and is called an antibracket.  When it is desired to emphasize the difference between brackets 
of two even variables and antibrackets, these will be written $[f,g]^{-}$ and $[\theta,\pi]^{+}$, 
respectively.  Between an odd and an even variable,
\begin{eqnarray}
[\theta,f]=\left \{ {\partial \theta \over \partial q_{i}} {\partial f \over \partial p^{i}} -
{\partial f \over \partial q_{i}} {\partial \theta \over \partial p^{i}} \right \} -
\left \{ {\partial \theta \over \partial \xi_{\alpha}} {\partial f \over \partial \pi^{\alpha}} +
{\partial f \over \partial \xi_{\alpha}} {\partial \theta \over \partial \pi^{\alpha}} \right \} \\
=-[f,\theta]. \nonumber
\end{eqnarray}
With these definitions the brackets form a (Grassmann) ring.  

Casalbuoni~\cite{Casalbuoni1976} 
shows that the set of Poisson brackets and antibrackets in pseudomechanics comprises a 
graded Lie algebra.  A graded Lie algebra is a Lie algebra containing both symmetric and 
antisymmetric bracket relations.~\cite{BerezinKac1970}$^{,}$~\cite{Corwin1975}  
Thus, in a quantum field theory which contains bosons and fermions and respects the symmetries 
of a Lie group, the set of field commutators and anticommutators make up a graded Lie algebra.  
The dynamical variables are "graded" by a degree that labels the symmetry of their brackets, e. g.:
\begin{eqnarray}
\delta_{q_{i}}=\delta_{p^{i}}=0 \\
\delta_{\xi_{\alpha}}=\delta_{\pi^{\alpha}}=1, \nonumber
\end{eqnarray}
where $\delta_{\sigma}$ is degree($\sigma$). The degree of the product of dynamical variables 
is the sum of their respective degrees modulo(2), so that
\begin{equation}
\delta_{\xi_{\alpha}}+\delta_{\pi^{\alpha}}=0.
\end{equation}
Thus, in order for the free Lagrangian for an odd dynamical variable to be an even quantity, 
it must be of the form 
\begin{equation}
L=i\pi\xi.
\end{equation}
The equation of motion of an anticommuting dynamical variable must therefore be first 
order.~\cite{Casalbuoni1976}  

The generalized Jacobi identity~\cite{Casalbuoni1976}$^{,}$~\cite{Weinberg2000} 
\begin{equation}
(-1)^{\delta_{\rho}\delta_{\pi}}[\gamma,[\rho,\pi]]+
(-1)^{\delta_{\rho}\delta_{\gamma}}[\rho,[\gamma,\pi]]+
(-1)^{\delta_{\gamma}\delta_{\pi}}[\pi,[\rho,\gamma]]=0 \label{eq:Jaobi}
\end{equation}
finds use in the derivation of canonical angular momentum ladder operators in Section \ref{ang. p}.

\subsection{Pseudoclassical Lagrangians} 

Classical Lagrangian theories of anticommuting Grassmann variables have been studied 
by Berezin and Marinov,~\cite{BM1977} Galvao and 
Teitelboim,~\cite{GalvaoTeitelboim1980} 
Casalbuoni,~\cite{Casalbuoni1976}$^{,}$~\cite{Casalbuoni1976a}   
Gomis \emph{et al.},~\cite{Gomis1985}$^{-}$~\cite{Gomis1986a} and others.  
In these investigations, the goal was to devise a supersymmetric model of classical point 
particles suitable for path integral quantization.  These models possess three kinds of symmetry:  
Poincar\'{e} invariance, which preserves the distinction between even and odd Grassman 
variables; supersymmetry under transformations which do not respect that distinction, 
instead relating even and odd Grassmann variables; and invariance under arbitrary monotone 
reparameterizations of the proper time.  This last is a form of gauge invariance.  
Constructing particle solutions with a well-defined world line that satisfies all these 
symmetry requirements turns out to be quite involved, especially the constraint analysis.  
Since we are unconcerned with quantization, we limit our attention to Poisson brackets.  
The construction of Dirac brackets described in Gomis \emph{et al.} and others 
receives no further discussion, other than to note that Dirac brackets 
necessarily have the same symmetry under exchange of arguments as do the Poisson brackets 
from which they are computed.

Of the three symmetries listed, only Poincar\'{e} invariance is of concern for the present 
discussion.  The simpler Lagrangian given by Di Vecchia and Ravndal~\cite{DiVecchia1979} 
and Ravndal~\cite{Ravndal1980} will serve as an illustrative example for the subsequent 
discussion, although the method of proof given in Section \ref{SST} applies to the more involved
models just mentioned as well.  Like those 
models, the Di Vecchia and Ravndal Lagrangian is invariant under 
supersymmetric transformations, but it differs from them in having a simple 
parameterization of proper time along a particle trajectory.  It is
\begin{equation}
L=\frac{1}{4}[\dot{q}^{\mu}\dot{q}_{\mu}-i\xi^{\mu}\dot{\xi}_{\mu}]. \label{eq:Ravndal}
\end{equation}
Here $q_{\mu}$ is an even position variable, and the intrinsic spin tensor for spin 1/2 is 
constructed from the spatial components of the Grassmann variable $\xi_{\mu}$.  The higher-spin 
generalization of eqn (\ref{eq:Ravndal}) is described below.   An overdot denotes the total 
differentiation with respect to a timelike affine parameter along a particle trajectory; 
in this model, the affine parameter is just proper time. 

The canonical momenta for free motion are 
\begin{equation}
p_{\mu}=\frac{1}{2}\dot{q}_{\mu}
\end{equation}
\begin{equation}
\pi_{\mu}=\frac{1}{4} i \xi_{\mu}.
\end{equation}
The free particle Hamiltonian is constant:
\begin{equation}
H=p^{2}=-m^{2}.
\end{equation}
In the presence of an electromagnetic field, the canonical momentum conjugate to $\xi_{\mu}$ 
is given by minimal coupling to a vector potential in the usual manner. 
In order to construct a classical theory for spin 1/2 which, upon quantization, yields the 
Dirac equation, the pseudomechanical models usually introduce a further odd Grassmann 
variable $\xi_{5}$, generalized suitably as needed for higher spin.~\cite{Gomis1986a}  
The Di Vecchia and Ravndal Lagrangian does not include this extra dynamical variable.

While it is necessary for Poincar\'{e} invariance that the $\xi_{\mu}$ be elements of a 
four-vector, the models must also impose $\xi_{0}$ =0 in some manner if 
the components of $\xi_{\mu}$ are to form an irreducible realization of the angular momentum 
algebra.  Berezin and Marinov,~\cite{BM1977} for example, posit an additional symmetry relation 
that results in the exclusion of $\xi_{0}$ from the equations of motion. 

\section{Irreducible Canonical Realizations of the Poincar\'{e} Group}

In quantum field theory, the notion of a particle is frequently identified with an 
irreducible unitary representation of the Poincar\'{e} 
group.~\cite{Weinberg1995}$^{,}$~\cite{Wigner1939}  The classical model of a massive particle 
used in this paper is the counterpart in the canonical formalism of such an irreducible unitary 
representation.~\cite{SMukunda1974}  
One speaks, instead, of irreducible canonical realizations, and replaces commutation relations 
amongst the matrix generators of infinitesimal Lorentz transformations and translations with 
Poisson brackets relating infinitesimal generators of canonical transformations.
Similarly, a function on phase space which is a function solely of the generators of the Lie algebra and 
which is an invariant in all realizations of the Lie group is called a Casimir invariant.  
Casimir invariants serve as the canonical equivalents of quantum-mechanical Casimir operators.
 
\subsection{The Poincar\'{e} Group}

The Lie algebra of the Poincar\'{e} group has a ten-parameter set of 
infinitesimal generators $\mathcal{M}$ and $\mathcal{P}$ satisfying commutation 
relations:~\cite{SMukunda1974}$^{,}$~\cite{Weinberg1995}$^{,}$~\cite{Dirac1949}$^{,}$~\cite{Weinberg1972}
\begin{eqnarray}
\mathcal{M}^{\mu\nu}\mathcal{M}^{\alpha\beta}-
\mathcal{M}^{\alpha\beta}\mathcal{M}^{\mu\nu}=
\eta^{\nu\beta}\mathcal{M}^{\mu\alpha}-\eta^{\nu\alpha}\mathcal{M}^{\mu\beta}-
\eta^{\mu\beta}\mathcal{M}^{\nu\alpha}
+\eta^{\mu\alpha}\mathcal{M}^{\nu\beta} \nonumber \\
\mathcal{M}^{\mu\nu}\mathcal{P}_{\alpha}-\mathcal{P}_{\alpha}\mathcal{M}^{\mu\nu}
=-\delta^{\nu}_{\alpha}\mathcal{P}^{\mu}+\delta^{\mu}_{\alpha}\mathcal{P}^{\nu} \\
\mathcal{P}_{\mu}\mathcal{P}_{\nu}-\mathcal{P}_{\nu}\mathcal{P}_{\mu}=0. \nonumber
\end{eqnarray}
Here $\eta$ is the Minkowski metric introduced earlier and $\mu,\nu$ range from 0-3.  The 
derivation of these relations is sketched in 
Appendix {\ref{appendixA}.
They give the Lie algebra of the generators of infinitesimal inhomogeneous Lorentz 
transformations near the origin.  Discrete transformations not deformable to the identity are 
described below.  $\mathcal{M}_{\mu\nu}$ is the generator of rotations in the $\mu-\nu$ plane, 
while $\mathcal{P}_{\mu}$ similarly generates spacetime translations in the $\mu$-direction.  
In Appendix {\ref{appendixB} it is shown that to each commutator of generators 
$\mathcal{A}_{s}$ of a representation of a Lie group, 
\begin{equation}
\mathcal{A}_{r}\mathcal{A}_{s}-\mathcal{A}_{s}\mathcal{A}_{r}=C^{t}_{rs}\mathcal{A}_{t}
\end{equation}
corresponds the Poisson bracket of generators of the equivalent canonical transformation
\begin{equation}
[A_{r},A_{s}]^{-}=C_{rs}^{t}A_{t}+d_{rs} \label{eq:LiePB}
\end{equation}  
where the $d$-matrices are constants.  In the case of the Poincar\'{e} group (but \emph{not} the 
Galilei group), it is possible to 
define the canonical generators $M_{\mu\nu}$ and $P_{\mu}$ equivalent to $\mathcal{M}_{\mu\nu}$ 
and $\mathcal{P}_{\mu}$ in such a way that the $d$-matrices vanish 
identically,~\cite{PauriProsperi1975} thus
\begin{equation}
[A_{r},A_{s}]^{-}=C_{rs}^{t} A_{t}.
\end{equation}
The Poisson bracket relations for the canonical realization of the Poincar\'{e} group are 
therefore
\begin{eqnarray}
\left [M^{\mu\nu},M^{\alpha\beta} \right ]^{-}=
\eta^{\nu\beta}M^{\mu\alpha}-\eta^{\nu\alpha}M^{\mu\beta}-\eta^{\mu\beta}M^{\nu\alpha}
+\eta^{\mu\alpha}M^{\nu\beta} \nonumber \\
\left [M^{\mu\nu},P_{\alpha} \right ]^{-}
=-P^{\mu}\delta^{\nu}_{\alpha}+P^{\nu}\delta^{\mu}_{\alpha} \\
\left [P_{\mu},P_{\nu} \right ]^{-}=0. \nonumber
\end{eqnarray}

The action of the ten generators of infinitesimal canonical transformations can be grouped as 
three spatial translations, one temporal translation, three boosts, and three 
rotations.~\cite{Weinberg1995}$^{,}$~\cite{PauriProsperi1975}  
Of these, only the bracket relations for the generators of rotational canonical transformations 
\begin{equation}
J_{i}=\epsilon_{ijk}M^{jk};
\end{equation}
\begin{equation}
[J_{i},J_{j}]^{-}=\epsilon_{ijk}J_{k}.
\end{equation}
($i,j,k$=1-3) find use in what follows.

\subsection{Angular momentum, spin, and irreducible canonical realizations in pseudomechanics}
\label{ang. p}

A (unitary) representation of a group is a set of linear transformations induced by a set 
of (unitary) matrices that gives a realization of the group; \emph{i.e.}, the matrix commutators 
are isomorphic to the group bracket relations.  In a canonical realization, the commutators 
are replaced by the equivalent Poisson brackets.  The discussion can be limited to 
irreducible realizations without loss of generality.  Within an irreducible realization of 
a Lie group, any two points of phase space can be connected by a canonical transformation 
representing the action of some element of the group.  An irreducible realization possesses 
no nontrivial invariants.  Thus, Casimir invariants reduce to numbers in an irreducible 
canonical realization. 

The properties of irreducible canonical realizations with definite angular momentum is 
obtained in a manner closely analogous to the corresponding quantum mechanical theory.  
Consider the transformation properties of dynamical variables in the canonical formalism 
under rotations.  The bracket relations obeyed by canonical angular momentum variables 
form a subgroup of the Poincar\'{e} group decoupled from the boost degrees of freedom.  
In particular, the spin degrees of freedom of a massive particle in the rest frame are 
treated exactly as in the nonrelativistic case.  Fix the direction of the $z$-axis along 
the spatial part of $\xi_{\mu}$ and write $\xi$ for its magnitude.  The infinitesimal canonical 
transformation induced by the generator of rotations about the $z$-axis in a dynamical variable is
\begin{equation}
\xi\Rightarrow\xi+\delta\phi[\xi,J_{z}].
\end{equation}
If the variable $\xi$ is rotationally symmetric about the axis defining $\phi$, the effect 
of this transformation must be equivalent to multiplication by a phase:
\begin{equation}
\xi\Rightarrow\xi+im\delta\phi\xi,
\end{equation}
or
\begin{equation}
[J_{z},\xi]=-im\xi. \label{eq:Jzbracket}
\end{equation}
This relation amounts to a kind of eigenvector condition.~\cite{Loinger1963}
Define the ladder operators
\begin{equation}
J_{\pm}=J_{z} \pm iJ_{y}.
\end{equation}
These have the following properties, closely analogous to the familiar quantum-mechanical 
identities
\begin{equation}
[J_{+},J_{-}]^{-}=-2iJ_{z}
\end{equation}
\begin{equation}
[J_{z},J_{\pm}]^{-}=\mp J_{\pm}. \label{eq:JzJpmbracket}
\end{equation}
Note that these relations are obtained from the quantum definitions by setting $i\hbar\equiv 1$.

It easily verified that the quantity
\begin{eqnarray}
J^{2}=J_{x}^{2}+J_{y}^{2}+J_{z}^{2} \\
=J_{z}^{2}+\frac{1}{2}[J_{+}J_{-}+J_{-}J_{+}]
\end{eqnarray}
has vanishing brackets with all the generators of rotations in an irreducible realization.  
It is thus a Casimir invariant which, in any irreducible canonical realization, is a constant 
number,~\cite{SMukunda1974b} so that~\cite{footnote3}
\begin{equation}
[J^{2},\xi]=const.\xi\equiv j(j+1)\xi.
\end{equation}

The irreducible realizations are labeled by the value of $j$.  In the remainder of this section, 
and the next, it is convenient to label $\xi$ by both eigenvalues $j$ and $m$ as $\xi_{jm}$, just 
as for irreducible tensor operators in spherical coordinates.  

Now consider the quantity
\begin{equation}
[J_{\pm},\xi_{jm}].
\end{equation}
We can determine the z-projection of its angular momentum by computing
\begin{equation}
[J_{z},[J_{\pm},\xi_{jm}]]
\end{equation}
with the aid of the Jacobi identity eqn (\ref{eq:Jaobi})
and the bracket relations of the ladder operators from eqn (\ref{eq:JzJpmbracket}):
\begin{equation}
[J_{z},[J_{\pm},\xi_{jm}]]=-im[J_{\pm},\xi_{jm}] \mp[J_{\pm},\xi_{jm}]
\end{equation}
or
\begin{equation}
[J_{z},[J_{\pm},\xi_{jm}]]=-i(m\pm 1)[J_{\pm},\xi_{jm}].
\end{equation}
Comparing this expression with eqn (\ref{eq:Jzbracket}) shows the result of taking the Poisson 
bracket of one of the ladder operators with a dynamical variable of spin $j$ is a dynamical 
variable with the $z$-projection $m$ of its dimensionless angular momentum changed by unity:
\begin{equation}
[J_{\pm},\xi_{jm}]=\lambda\xi_{jm\pm 1}.
\end{equation}
where $\lambda$ will depend on $j$ and $m$.

Proceeding in this vein, the action of the ladder operators in the canonical formalism is 
obtained in close analogy to the quantum case.  In particular,~\cite{Loinger1963}
\begin{equation}
[J_{\pm},\xi_{jm}]=-i\sqrt{(j\mp m)(j\pm m+1)}\xi_{jm\pm1}. \label{eq:ladderopdef}
\end{equation}
In a manner entirely analogous to the quantum 
case~\cite{Tung1985a}$^{-}$~\cite{Loinger1963}$^{,}$~\cite{Dirac1958}
it can be shown that the eigenvectors of 
$J_{z}$ given by eqn (\ref{eq:Jzbracket}) have integer eigenvalues 
\begin{equation}
-j\leq m \leq j
\end{equation}
and that they span a $2j+1$ dimensional subspace of the Hilbert space of canonical realizations of 
the rotation subgroup.  It is the behavior one expects, say, from a spherical harmonic of angular 
momentum $j$, with $z$-projection $m$.  Poisson brackets being 
dimensionless, the quantities $j$ and $m$ that label distinct elements of an irreducible
canonical realization with definite angular momentum do not set a scale for the physical 
angular momentum associated with the dynamical variable $\xi$.  The machinery developed in this
section may, therefore, have the appearance of a mathematical analogy devoid of physical content, 
but it is required in order to obtain the properties of classical Grassman dynamical variables
under time inversion in Section \ref{T}.

The canonical formalism just sketched accommodates intrinsic spin, in much the same manner as its
quantum-mechanical counterpart.  The infinitesimal canonical generator of rotations $M^{\mu\nu}$ 
for point particle motion can be written 
\begin{equation}
M^{\mu\nu}=x^{\mu}p^{\nu}-x^{\nu}p^{\mu}+S^{\mu\nu}. \label{eq:Mmunudef}
\end{equation}
$S^{\mu\nu}$, of course, is the classical intrinsic spin tensor of the particle.  This result 
may be obtained with aid of the classical Pauli-Lubanski vector~\cite{SMukunda1974c} 
or directly from even~\cite{PauriProsperi1975} and odd~\cite{Gomis1985} 
irreducible canonical realizations. 
From eqn (\ref{eq:Mmunudef}), the intrinsic spin vector of a particle in its rest frame is 
obtained as
\begin{equation}
S_{i}=\epsilon_{ijk}S^{jk}
\end{equation}
with
\begin{equation}
[S_{i},S_{j}]^{-}=\epsilon_{ijk}S_{k}.
\end{equation}
In any frame in which the momentum $\mathbf{p}$ vanishes, $\mathbf{J}=\bf{S}$.  Pauri and 
Prosperi~\cite{PauriProsperi1967} construct the irreducible canonical realizations of spin $j$, 
and note that they 
supply the classical equivalent of a particle with spin.  Gomis \emph{et al.}\cite{Gomis1986a} 
extend the construction to arbitrary spin Grassmann variables with the canonical version of the 
Bargmann-Wigner formalism.~\cite{BargmannWigner1948}$^{,}$~\cite{FierzPauli1939}  
The corresponding generalization of the Di Vecchia and Ravndal model spin tensor in terms 
of a set of spin 1/2 Grassman variables $\xi_{\lambda}^{\mu}$ is given by
\begin{equation}
S^{\mu\nu}=
-\frac{i}{2}\sum_{\lambda=1}^{N}\xi^{\mu}_{\lambda}\xi^{\nu}_{\lambda}, \label{eq:sptensor}
\end{equation}
where the spin is N/2.  The spin portion of the Lagrangian becomes
\begin{eqnarray}
L_{spin}=
-\frac{i}{4}\eta_{\mu\nu}\sum_{\lambda=1}^{N}\xi_{\lambda}^{\mu}\dot{\xi}_{\lambda}^{\nu} \\
\equiv-\frac{i}{4}\Theta^{\mu}\cdot\dot{\Theta}_{\mu}, \label{eq:Lspin}
\end{eqnarray}  
by way of defining both the dynamical variable $\Theta$ 
for spin N/2 and the scalar product on spin indices.  The form of the spin tensor 
in eqn (\ref{eq:sptensor}) creates problems when either spacetime 
index $\mu$ or $\nu$ is 0.~\cite{BM1977}  In the Di Vecchia and Ravndal Lagrangian, 
the spin angular momentum tensor satisfies
\begin{equation}
p_{\mu}S^{\mu\nu}=0
\end{equation}
identically, so that the spin has only spatial components in the particle rest 
frame.~\cite{Ravndal1980}

\section{Time reversal, weak and strong, in pseudomechanics} \label{T}

\subsection{Time reversal invariance and anticanonical transformations}

The preceding Section developed the properties of continuous coordinate transformations upon 
classical Grassmann variables, specifically rotations, necessary for proving the 
spin-statistics connection.  A complete realization of the  Poincar\'{e} 
group for massive spinning particles must also include the discrete transformations of parity, 
time reversal, and charge conjugation.  The effects
of parity and time reversal are given by 
\begin{equation}
\mathcal{\Pi}(\phi(\mathbf{x},t)=\phi(-\mathbf{x},t)
\end{equation}
\begin{equation}
\mathcal{T}(\phi(\mathbf{x},t)=\phi(\mathbf{x},-t)
\end{equation}
The full Poincar\'{e} group thus has four components related by the various combinations of the 
parity and time-reversal transformations.  The classical analog of charge conjugation is discussed 
in Section \ref{C}.  Schwinger's proof of the spin-statistics connection 
depends upon the effects of time-reversal and charge-conjugation on states of definite spin.  The 
parity operation is of no further concern here.~\cite{footnote4} 

The operation $\mathcal{T}$ commutes with the generators of spatial translations and rotations, but anticommutes with 
the generators of boosts and, in particular, of time translations:
\begin{equation}
\mathcal{T}\mathcal{P}^{0}=-\mathcal{P}^{0}\mathcal{T} \label{eq:TPIcommutator}
\end{equation}
Let $\hat{T}$ be the operator which realizes $\mathcal{T}$ on functions in phase space $(q,p)$
\begin{equation}
\hat{T}(\phi(q,p))=\phi(q',p')
\end{equation}
where the primed variables are related to the unprimed ones by a canonical transformation
\begin{equation}
q'=q'(q,p)
\end{equation}
\begin{equation}
p'=p'(q,p).
\end{equation}
We have, from eqn (\ref{eq:TPIcommutator}),
\begin{equation}
\hat{T}([E,\phi])=-[E,\phi] \hat{T}, \forall \phi,
\end{equation}
while from the definition of a canonical transformation 
\begin{equation}
[\phi,\gamma]_{qp}=[\phi,\gamma]_{q'p'}
\end{equation}
we have
\begin{equation}
\hat{T}[\phi,\gamma]=[\hat{T}(\phi),\hat{T}(\gamma)],
\end{equation}
so that
\begin{equation}
([\hat{T}(E),\hat{T}(\phi)])=-[E,\hat{T}(\phi)]
\end{equation}
for arbitrary $\phi$. $\hat{T}(E)$ and -E can therefore differ by only a constant.  As 
$\hat{T}^{2}$ must equal unity,
\begin{equation}
\hat{T}(E)=-E.
\end{equation}
But this is awkward, because the generator of time translations is interpreted as the 
energy, and should be positive definite.  The solution~\cite{PauriProsperi1975} is to 
realize time reversal $\mathcal{T}$ as an anticanonical operation $T$,
\begin{equation}
T([\phi,\gamma])=-[T(\phi),T(\gamma)] \label{eq:anticanonicality}
\end{equation}
with
\begin{equation}
T(E)=E.
\end{equation}
The anticanonical time reversal operation commutes with the generator of boosts and 
anticommutes with generators of rotation and translations.  In particular, 
\begin{equation}
T(\bf{J})=-\bf{J}.  \label{eq:TonJ}
\end{equation}

The quantum mechanical realization of time reversal is an antilinear and antiunitary 
transformation.  Wigner~\cite{Wigner1959} shows it is always possible to write such 
a transformation as the composition of a unitary transformation with complex conjugation.  
To maintain consistency with the quantum case-by way of inverting the correspondence 
principle-$T$ is defined as an antilinear, as well as anticanonical, operation
\begin{equation}
T(a\phi+b\gamma)=a^{*}T(\phi)+b^{*}T(\gamma).
\end{equation}
An antilinear operation is likewise the composition of complex conjugation with a linear 
transformation.  Note that the action of $T$, as defined, on a scalar quantity 
is that of complex conjugation.  This observation finds use in Section \ref{WTprods} for 
finding the effect of time inversion on products of dynamical variables.  

\subsection{Weak Time-reversal symmetry in pseudomechanics}  

We next address the effect of time inversion on the angular momentum relations given earlier, 
in particular the raising and lowering operations eqn (\ref{eq:ladderopdef}).~\cite{Weinberg1995b}  
First, consider the effect of $T$ on eqn (\ref{eq:Jzbracket})
\begin{equation}
T(\xi_{jm}) \Rightarrow T(\xi_{jm})-im\delta\phi T(\xi_{jm}),
\end{equation}
or
\begin{equation}
[J_{z},T(\xi_{jm})]=imT(\xi_{jm}),
\end{equation}
from which
\begin{equation}
T(\xi_{jm})=\varsigma_{m}\xi_{j-m}
\end{equation}
where $\varsigma_{m}$ may depend on $j$ as well.  Second, the anticanonical action of $T$ gives
\begin{equation}
T([J_{\pm},\xi_{jm}])=-[T(J_{\pm}),T(\xi_{jm})].
\end{equation}
Now, recalling eqns (\ref{eq:ladderopdef}), (\ref{eq:anticanonicality}) and (\ref{eq:TonJ})
\begin{eqnarray}
T([J_{\pm},\xi_{jm}])=i\sqrt{(j\mp m)(j\pm m+1)}\varsigma_{m\pm1}\xi_{j-(m\pm1)} 
\label{eq:TonbracketA} \\
=-[T(J_{\pm}),T(\xi_{jm})]=[J_{x}\mp iJ_{y},\varsigma_{m}\xi_{j-m}] \nonumber \\
=-i\sqrt{(j\mp m)(j\pm m+1)}\varsigma_{m}\xi_{j-(m\pm1)}. \label{eq:TonbracketB}
\end{eqnarray}
Dividing out common terms in eqns (\ref{eq:TonbracketA}) and (\ref{eq:TonbracketB}) gives
\begin{equation}
-\varsigma_{m}=\varsigma_{m+1}
\end{equation}
or 
\begin{equation}
\varsigma_{m}=\varsigma(-1)^{-m}.
\end{equation}
It remains to fix $\varsigma$.  The choice $\varsigma_{j}=1$ ensures that $T$ 
applied to a real dynamical variable will give a real result, giving
\begin{equation}
\varsigma_{m}=(-1)^{j-m}.
\end{equation}
Thus,
\begin{equation}
T(\xi_{jm})=(-1)^{j-m}\xi_{j-m}.
\end{equation}

\subsection{Weak T on products of dynamical variables} \label{WTprods}

The Lagrangian is constructed from invariant scalar combinations of dynamical variables 
and their derivatives.  In pseudomechanical models, the invariant used is the scalar 
product of two dynamical variables with the same spin.  The spin portion of the Lagrangian 
from eqn (\ref{eq:Lspin}) may be written
as a sum of scalar products of the form~\cite{Edmonds1960}$^{,}$~\cite{FanoRacah1959}
\begin{equation}
\sigma \cdot \pi=(-1)^{k}\sigma_{k}\pi_{-k}.
\end{equation}
Consider the action of $T$ on the 
scalar product of two dynamical variables belonging to the same irreducible canonical 
realization of spin $j$.  Recall that the antilinear action of T on a scalar quantity is that 
of complex conjugation 
and, from eqn (\ref{eq:cconj}), that complex conjugation inverts the order of factors in a 
product:
\begin{equation}
(\sigma\pi)^{*}=\pi^{*}\sigma^{*}
\end{equation}
Thus,
\begin{eqnarray}
T(\sigma \cdot \pi)=(-1)^{k}\pi_{-k}^{*}\sigma_{k}^{*}=(-1)^{-j}\pi_{k}^{*}(-1)^{j-k}
\sigma_{-k}^{*} \nonumber \\
=(-1)^{-j}\pi_{k}^{*}T(\sigma_{k})=(-1)^{-2j}(-1)^{k}T(\pi_{k})T(\sigma_{-k}) \\
=(-1)^{2j}T(\pi) \cdot T(\sigma). \nonumber 
\end{eqnarray}
The inversion of the order of factors under $T$ clearly generalizes by induction to an arbitrary 
number of them.

\subsection{Charge conjugation in pseudomechanics and strong time-reversal invariance} \label{C}

Schwinger's proof of the spin-statistics connection relies upon "strong", or "Weyl", time 
reversal, as opposed to the "weak" or "Wigner" time reversal $T$ as defined above.  The 
condition of strong time reversal invariance is that the form of the classical action be preserved 
if evolution from an initial to a final state is replaced by the evolution of a time-reversed 
state from the final state to the initial one.  That is, in addition to reversing the sign of 
the locally timelike variable in all dynamical quantities, initial and final states are 
exchanged in the action, and the affine parameter labeling proper time changes sign as well: 
\begin{equation}
\tau \Rightarrow -\tau
\end{equation}
\begin{equation}
{d \over d\tau} \Rightarrow -{d \over d\tau}
\end{equation}
Costella \emph{et al.}~\cite{Costella1997} show that this operation is the classical analog of 
charge conjugation.  
The connection with St\"{u}ckelberg's identification of antiparticle motion with time-reversed 
particle motion is clear; in fact, Feynman~\cite{Feynman1948} examined the classical formulation 
of this 
concept prior to the debut of his theory of positrons.~\cite{Feynman1949}  Strong time reversal 
is therefore the composition of the weak time reversal operation $T$ with charge conjugation 
$C$.~\cite{Schwinger1951}$^{,}$~\cite{Schwinger1958}$^{,}$~\cite{DuckS1997}$^{,}$~\cite{DuckS1998}

\section{Connection between spin and statistics} \label{SST}

With the results from preceding sections on properties of classical Grassmann variables of 
definite spin under time-reversal and charge-conjugation in hand, we are now in position 
to impose the condition of invariance under strong time-reversal transformation upon a 
pseudomechanical system and show that the spin-statistics connection necessarily 
follows.  Invariance of the pseudoclassical description of particle motion under strong $T$ 
inversion requires that the form of the action functional 
\begin{equation}
S=\int_{\tau_{1}}^{\tau_{2}} d\tau L(\theta,\dot{\theta})
\end{equation}
be unaltered by inverting both the sign of t and the definition of proper time, to include 
the order of the initial and final proper time of a segment of a particle orbit.  The 
former operation is $T$; the latter, $C$.  Thus, if
\begin{equation}
S=\int_{\tau_{1}}^{\tau_{2}} d\tau L(\theta,\dot{\theta}) \Rightarrow 
\int_{\tau_{2}}^{\tau_{1}} d\tau' L(T(\theta),{d \over d\tau'}T(\theta))
\end{equation}
under the combined operation of $T$ and $C$, the form of the action will be preserved.  The 
total Lagrangian in pseudoclassical models is made of quadratic forms in the dynamical 
variables.  It suffices to consider the contribution $L_{j}$ for an irreducible realization of 
spin j.~\cite{Gomisnote}  Apply the operations of $T$ and $C$ to $L_{j}$.  First $T$: 
\begin{eqnarray}
T(L_{j})=(-1)^{2j}L^{t}_{j}(T(\theta),{d \over d\tau}T(\theta)) \\
=(-1)^{2j}L^{t}_{j}(T(\theta),-{d \over d\tau'}T(\theta)) \nonumber
\end{eqnarray}
where the superscript $t$ on $L_{j}$ indicates transposition of the order of all factors in the 
Lagrangian.

Next, applying $C$ changes the sign of the proper time derivative:
\begin{equation}
CT(L_{j})=(-1)^{2j}L^{t}_{j}(T(\theta),{d \over d\tau'}T(\theta))
\end{equation}
Invariance of the form of the action under $CT$ is guaranteed if
\begin{equation}
L_{j}(T(\theta),{d \over d\tau'}T(\theta))=
(-1)^{2j}L^{t}_{j}(T(\theta),{d \over d\tau'}T(\theta)).
\end{equation}

This last will hold if the sign change attendant upon inverting the order of odd Grassmann 
variables is compensated by a factor of minus one in front, while no compensating sign 
change accompanies inverting the order of even Grassmann variables.  We conclude classical 
spin variables which are irreducible canonical realizations of 
spin $j$ must be commuting, even Grassmann variables if $j$ is an integer, and anticommuting, 
odd Grassmann variables if $j$ is half-integral.  From the symmetry properties of brackets 
in pseudomechanics given earlier follows immediately the conclusion that irreducible canonical 
realizations for integral $j$ obey Poisson bracket relations, while realizations for half-integral 
$j$ obey Poisson antibracket relations.  The Poisson brackets for the spin degrees of freedom 
are
\begin{equation}
[ \theta_{\mu},\theta_{\nu}]^{-}=[\pi_{\mu},\pi_{\nu} ]^{-}=0 \label{eq:SSTB}
\end{equation} 
\begin{equation}
[\theta_{\mu},\pi_{\nu}]^{-}=\eta_{\mu\nu} \nonumber
\end{equation}
for $2j$=even, and
\begin{equation}
[ \theta_{\mu},\theta_{\nu} ]^{+}=[\pi_{\mu},\pi_{\nu}]^{+}=0 \label{eq:SSTF}
\end{equation} 
\begin{equation}
[ \theta_{\mu},\pi_{\nu} ]^{+}=-\eta_{\mu\nu} \nonumber
\end{equation}
for $2j$=odd.  The vectorial position variables $q_{\mu}$ and $p_{\mu}$ satisfy Poisson bracket 
relations for any value of $j$.  This is the spin-statistics theorem stated in the language of the
canonical formalism for pseudomechanics. 

\section{Comments}

The result just obtained is neither the strongest, nor the most general, that could be desired.  
It is as close to a literal transcription of Schwinger's 1951 reasoning into the language of 
the canonical formalism for particle mechanics as could be contrived.  The choice to proceed 
in this manner was not made as a simple matter of filial piety.  Rather, it appears the 
simplest, quickest route to a classical spin statistics relation is to recapitulate Schwinger's 
proof in close to its original form.  It is notable that the proof proper in Section V is 
shorter, and arguably more appealing, than the formal developments necessary to erect the 
canonical formalism underpinnings which preceded it. 

Like that earlier proof, this one applies to free particles, or to particles minimally 
coupled at most weakly by an interaction that conserves $CT$.  This should not be considered a 
serious shortcoming in the case of the electromagnetic interaction, in which $C$, $P$, and $T$ are 
each conserved separately.  A classical spin-statistics connection valid for electromagnetically 
interacting particles would seem capable of meeting most needs for that class of 
relation.~\cite{footnote7}  

There is another, and more serious, sense in which the result just shown should be regarded 
as a comparatively weak one.  Schwinger's argument, strictly speaking, applies only to brackets 
constructed from dynamical variables evaluated at a common point of phase space.  The point 
might seem of limited relevance for eqns (\ref{eq:SSTB}) and (\ref{eq:SSTF}), since that is 
how one normally evaluates Poisson brackets, but the commutation properties of dynamical 
variables at distinct phase space locations is left undetermined by the present argument.  
Their extension even to separate points at null interval, which would be the minimum required for 
consistency with the relation just shown, does not follow without additional assumptions.  
Schwinger~\cite{Schwinger1951} cites "the general compatibility requirement for physical 
quantities attached to points with a spacelike interval" to justify extending commutation 
relations from coincident to spacelike intervals.  The assumptions of a particular graded Lie 
algebra structure for the bracket relations in the present discussion could similarly be 
strengthened by fiat, but only at the price of underscoring the weakness of the result obtained. 

The particular form of classical mechanics used in the foregoing may look odd as an exemplar 
of classical physics-amongst other peculiarities, dynamical variables are 
allowed to take on complex values, in general.  While it is often said that classical dynamical 
variables should be real-valued functions on spacetime, the classical physics of waves or 
oscillatory phenomena is too riddled with complex exponentials for this stricture to be 
altogether convincing.  In any event, complex conjugation is required in the present 
demonstration for, strictly speaking, its effect on real Grassman variables.  

What should not be obscured, however, is that pseudomechanics offers an elementary example 
of a physical theory which respects the spin-statistics connection without being quantum 
mechanical. 

\section{Conclusion}

"We conclude that the connection between spin and statistics of particles is implicit in 
the requirement of invariance under coordinate transformations."  Schwinger, 1951

Schwinger used this comment as a period for his proof of the spin-statistics connection.  
In spirit, it is very close to that of Pauli's, cited at the start of this paper, 
but in wording it is notably less emphatic.  Oddly so, given that the requirement of invariance 
upon which Schwinger erected his construction of quantum electrodynamics from the action 
principle was Poincar\'{e} invariance.  Note that in either statement the tone struck can be 
interpreted as a classical one.

It has been remarked more than 
once~\cite{DuckS1997}$^{,}$~\cite{Balach1992}$^{,}$~\cite{Balach1993}$^{,}$~\cite{DuckS1998} 
that proofs of the spin-statistics connection necessarily 
depend upon some assumption traceable to Poincar\'{e} invariance.  The dependence may be 
explicit, as in Pauli's original proof, or the axiomatic proofs of Burgoyne, L\"{u}ders, and 
Zumino, or it may be implicit, as in Feynman,~\cite{Feynman1987} the topological proof of 
Balachandran \emph{et al.}, or the proof of Berry and Robbins~\cite{BR1997} using topological 
phases.  The topological theorems~\cite{Balach1992}$^{,}$~\cite{Balach1993}$^{,}$~\cite{RF1968} 
invoke the existence of antiparticles.  Proofs by Weinberg,~\cite{Weinberg1964} which use the 
language of representations of the Poincar\'{e} group, or of Berry and Robbins, which do not, 
invoke no symmetry higher than that of rotational invariance.  But the Poincar\'{e} group contains 
rotational symmetry as a subgroup. 

The physical world is not more relativistic than it is quantum-mechanical. However, the 
existence of classical systems obeying the spin-statistics connection allows one to think of 
that phenomenon as a relativistic one at bottom.  Under terrestrial laboratory conditions, 
Poincar\'{e} invariance is an exquisitely accurate symmetry of nature.  The lesser symmetries 
upon which the spin-statistics connection depends, be they the existence of antiparticles, or of 
rotational invariance, or (as here) invariance under time inversion, are all necessary 
\textit{consequences} of Poincar\'{e} invariance in nature. 

It is usually supposed that relativistic phenomena are significant only for high energies, or 
for velocities approaching that of light.  The effect of the spin-statistics connection on the 
nature of the everyday world is profound, perhaps most significantly under intrinsically low 
energy conditions of Fermi degeneracy.  It cannot be doubted that, were the spin-statistics 
connection different, or nonexistent, the resulting world would almost certainly be 
unrecognizable to us.  To suppose such a violent rearrangement of microscopic physics would 
leave the macroscopic world sensibly unaltered amounts to invoking a conspiracy of nature for 
the sake of avoiding that conclusion.  Not one of our senses is independent of the accidents 
consequent upon the connection between spin and statistics, either from the role played by 
Pauli exclusion in atomic structure and chemical binding, or from the effects of incompressible 
flow; not sight, nor hearing, nor smell, nor taste, nor touch.  One need not invoke exotic 
conditions to find evidence of relativistic symmetries in the world.

\begin{acknowledgments}
I wish to thank Drs. S. Gasster, J. Johnson, and G. Smit for their careful reading of this paper, 
and for suggestions which subtantially improved its clarity.
\end{acknowledgments}

\appendix 

\section{The Lie algebra of the Poincar\'{e} Group} \label{appendixA}

The general inhomogeneous Lorentz transformation on a four-vector $x_{\mu}$ is
\begin{equation}
x^{\prime\mu}=\Lambda^{\mu}_{\alpha}x^{\alpha}+a^{\mu} \label{eq:inhomLorentz}
\end{equation}
If the four-interval between two points in spacetime is to remain invariant under 
eqn (\ref{eq:inhomLorentz}), then the $\Lambda$ matrices must satisfy
\begin{equation}
\Lambda^{\mu}_{\alpha}\Lambda^{\nu}_{\beta}\eta_{\mu\nu}=\eta_{\alpha\beta}. \label{eq:etainv}
\end{equation}
Let a set of (unitary) matrices D comprise a representation of inhomogeneous Lorentz 
transformations satisfying
\begin{equation}
D(\Lambda_{1})D(\Lambda_{2})=D(\Lambda_{1}\Lambda_{2}) \label{eq:homomorphism}
\end{equation}
and consider transformations infinitesimally close to the origin,
\begin{equation}
\Lambda^{\mu}_{\alpha}=\delta^{\mu}_{\alpha}+\omega^{\mu}_{\alpha}+O(\omega^{2}) \label{eq:Lx}
\end{equation}
and
\begin{equation}
a^{\mu}=\epsilon^{\mu} \label{eq:Px}
\end{equation}
with $|\omega|,|\epsilon| \ll 1.$
If Eqn (\ref{eq:etainv}) is to be satisfied, $\omega$ must be antisymmetric in its indices.  The 
corresponding matrix representaton of eqns (\ref{eq:Lx}) and (\ref{eq:Px}) is
\begin{equation}
D(1+\omega,\epsilon)=1+\frac{1}{2}\omega_{\mu\nu}\mathcal{M}^{\mu\nu}-\epsilon_{\rho}
\mathcal{P}^{\rho}
\end{equation}
to first order, where $\mathbf{\mathcal{M}}$ is a constant antisymmetric matrix and 
$\mathbf{\mathcal{P}}$ is a vector.  In order for eqn 
(\ref{eq:homomorphism}) to be satisfied, $\mathbf{\mathcal{M}}$ and $\mathbf{\mathcal{P}}$ 
must 
obey~\cite{SMukunda1974}$^{,}$~\cite{Weinberg1995}$^{,}$~\cite{Dirac1949}$^{,}$~\cite{Weinberg1972}
\begin{eqnarray}
\mathcal{M}^{\mu\nu}\mathcal{M}^{\alpha\beta}-
\mathcal{M}^{\alpha\beta}\mathcal{M}^{\mu\nu}=
\eta^{\nu\beta}\mathcal{M}^{\mu\alpha}-\eta^{\nu\alpha}\mathcal{M}^{\mu\beta}-
\eta^{\mu\beta}\mathcal{M}^{\nu\alpha}
+\eta^{\mu\alpha}\mathcal{M}^{\nu\beta} \nonumber \\
\mathcal{M}^{\mu\nu}\mathcal{P}_{\alpha}-\mathcal{P}_{\alpha}\mathcal{M}^{\mu\nu}
=-\delta^{\nu}_{\alpha}\mathcal{P}^{\mu}+\delta^{\mu}_{\alpha}\mathcal{P}^{\nu} \\
\mathcal{P}_{\mu}\mathcal{P}_{\nu}-\mathcal{P}_{\nu}\mathcal{P}_{\mu}=0. \nonumber
\end{eqnarray}

\section{Canonical Realizations of a Lie Group} \label{appendixB}

The group action of a symmetry in the canonical formalism leads to a corresponding Lie algebra of 
Poisson brackets.  Let the commutation relations of the infinitesimal generators of a Lie group 
be written
\begin{equation}
\mathcal{A}_{r}\mathcal{A}_{s}-\mathcal{A}_{s}\mathcal{A}_{r}=C^{t}_{rs}\mathcal{A}_{t},
\label{eq:Liebracket}
\end{equation}
and let the canonical coordinates and momenta, $(q_{\mu},p_{\nu})$ even and $(\xi_{\mu},
\pi_{\nu})$ odd, obey fundamental Poisson bracket relations
\begin{eqnarray}
\left [q_{\mu},p_{\nu} \right ]=\eta_{\mu\nu} \nonumber \\
\left [ \xi_{\mu},\pi_{\nu} \right ]=-\eta_{\mu\nu} \label{eq:fundPBs} \\
\left [q_{\mu},q_{\nu} \right ]=\left [p_{\mu},p_{\nu} \right ]
= \left [\xi_{\mu},\xi_{\nu} \right ]= \left [\pi_{\mu},\pi_{\nu} \right ]=0. \nonumber
\end{eqnarray}
A canonical realization of a symmetry group is a set of transformations of the canonical 
coordinates, homomorphic to the symmetry group, that leaves the fundamental bracket relations 
eqns (\ref{eq:fundPBs}) unaltered.  The infinitesimal canonical transformations are defined 
so as not to mix even and odd Grassmann variables.~\cite{Casalbuoni1976a}  That is, 
\begin{equation}
q'_{i}=q'_{i}(\{q\},\{p\},\{a\})
\end{equation}
\begin{equation}
p'_{i}=p'_{i}(\{q\},\{p\},\{a\})
\end{equation}
where each of the set of parameters $\{a\}$ which characterizes the transformation is even, 
and
\begin{equation}
\theta'_{\alpha}=\theta'_{\alpha}(\{\theta\},\{\pi\},\{\rho\})
\end{equation}
\begin{equation}
\pi'_{\alpha}=\pi'_{\alpha}(\{\theta\},\{\pi\},\{\rho\})
\end{equation}
where the set of $\{\rho\}'s$ is odd.  With canonical transformations so restricted, the 
generators of infinitesimal canonical transformations $A_{r}$ are even functions of the 
canonical variables.  Then
\begin{eqnarray}
q'_{i}=q_{i}+\delta a^{r} \left [A_{r},q_{i} \right ]^{-} \label{eq:infCTA} \\
p'_{i}=p_{i}+\delta a^{r} \left [A_{r},p_{i} \right ]^{-} \nonumber
\end{eqnarray}
and
\begin{eqnarray}
\theta'_{\alpha}=
\theta_{\alpha}+\delta a^{s} \left [A_{s},\theta_{\alpha} \right ]^{-} \label{eq:infCTB} \\
\pi'_{\alpha}=\pi_{\alpha}+\delta a^{s} \left [A_{s},\pi_{\alpha} \right ]^{-} \nonumber
\end{eqnarray}
correspond to the infinitesimal operations
\begin{equation}
1+\delta a^{r} \mathcal{A}_{r} \label{eq:infxformA}
\end{equation}
and
\begin{equation}
1+\delta \rho^{s} \mathcal{A}_{s}, \label{eq:infxformB}
\end{equation}
respectively.  These are the canonical generators of infinitesimal transformations, and 
their bracket relations must form a realization of the Lie algebra.  In particular, to 
the infinitesimal transformation
\begin{equation}
1+\delta a^{r} \delta b^{s} (\mathcal{A}_{r}\mathcal{A}_{s}-\mathcal{A}_{s}\mathcal{A}_{r})
\label{eq:secondorderA}
\end{equation}
must correspond the canonical transformation
\begin{equation}
q'_{i}=q_{i}+\delta a^{r} \delta b^{s} \left [\left [A_{r},A_{s}\right]^{-},q_{i} \right ]
\end{equation}
\begin{equation}
p'_{i}=p_{i}+\delta a^{r} \delta b^{s} \left [\left [A_{r},A_{s}\right]^{-},p_{i} \right ]
\end{equation}
and to
\begin{equation}
1+\delta \rho^{t} \delta \upsilon^{u} 
(\mathcal{A}_{t}\mathcal{A}_{u}-\mathcal{A}_{u}\mathcal{A}_{t}),
\label{eq:secondorderB}
\end{equation}
\begin{equation}
\theta'_{\alpha}=\theta_{\alpha}+\delta \rho^{t} \delta \upsilon^{u} 
\left [\left [A_{t},A_{u}\right]^{-},\theta_{\alpha} \right ]
\end{equation}
\begin{equation}
\pi'_{\alpha}=\pi_{\alpha}+\delta \rho^{t} \delta \upsilon^{u}
 \left [\left [A_{t},A_{u}\right]^{-},\pi_{\alpha} \right ].
\end{equation}
Inserting eqn (\ref{eq:Liebracket}) into eqns (\ref{eq:secondorderA}) and 
(\ref{eq:secondorderB}), and comparing eqn (\ref{eq:infxformA}) with eqn (\ref{eq:infCTA}), 
and eqn (\ref{eq:infxformB}) with eqn (\ref{eq:infCTB}), leads to the conclusion 
that~\cite{PauriProsperi1966}
\begin{equation}
[A_{r},A_{s}]^{-}=C_{rs}^{t}A_{t}+d_{rs}
\end{equation}
where the d's are constants.


\begin{thebibliography}{5}
\bibitem{footnote2} It may be that electon-hole pairs would suffice in condensed matter 
applications.
\bibitem{Boltznote}The result to be proved concerns symmetries of Poisson brackets for dynamical 
variables of a single massive particle; there is no need to examine what is meant by "identical" 
particles in the present discussion.
\bibitem{footnote4} Strictly speaking, the discrete transformation $CT$ used later is a 
surrogate for $P$ by virtue of the $CPT$ theorem.  However, the discussion here is restricted 
in such a way that $CPT$ need not be explicitly invoked.  The components connected by $T$ make up 
the orthochorous Poincar\'{e} group.
\bibitem{footnote7} Schwinger's original proof can be generalized by substituting flavor 
symmetry for $CT$ invariance.
\end{thebibliography}
\end{document}